\documentclass[11pt]{article}

\usepackage[utf8]{inputenc}
\usepackage[T1]{fontenc}
\usepackage{lmodern}
\usepackage[margin=1in]{geometry}
\usepackage{amsmath,amssymb}
\usepackage{graphicx}
\usepackage{booktabs}
\usepackage{hyperref}
\usepackage{xcolor}
\usepackage{listings}
\usepackage{natbib}
\usepackage{float}
\usepackage{placeins}  % \FloatBarrier keeps single-column figures inside their (sub)section
\usepackage{enumitem}
\usepackage{url}
\usepackage{microtype}
\usepackage{multirow}
\usepackage{tikz}
\usetikzlibrary{positioning,arrows.meta,fit,calc,backgrounds,shapes.geometric}
\tikzset{
  io/.style={draw,rounded corners,fill=black!5,inner sep=3pt,align=center,minimum height=7mm,font=\footnotesize},
  llm/.style={draw,rounded corners,fill=blue!10,inner sep=3pt,align=center,minimum height=7mm,font=\footnotesize},
  gr/.style={draw,rounded corners,fill=green!14,inner sep=3pt,align=center,minimum height=7mm,font=\footnotesize},
  doc/.style={draw,rounded corners,fill=orange!16,align=center,inner sep=3pt,minimum height=7mm,font=\footnotesize},
  dec/.style={draw,diamond,aspect=2,fill=yellow!18,inner sep=1pt,align=center,font=\scriptsize},
  fl/.style={-{Latex},semithick},
  fld/.style={-{Latex},semithick,dashed},
}

\hypersetup{
  colorlinks=true,
  linkcolor=blue!70!black,
  citecolor=blue!70!black,
  urlcolor=blue!70!black
}

\lstdefinelanguage{Cypher}{
  keywords={MATCH, RETURN, WHERE, CREATE, DELETE, SET, REMOVE, MERGE, WITH, UNWIND, UNION, ORDER, BY, SKIP, LIMIT, EXPLAIN, PROFILE, CALL, YIELD, AS, DISTINCT, OPTIONAL, AND, OR, NOT, IN, IS, NULL, TRUE, FALSE, EXISTS, CASE, WHEN, THEN, ELSE, END, DROP, SHOW, INDEX, INDEXES, CONSTRAINTS, ON},
  sensitive=false,
  morestring=[b]',
  morestring=[b]",
  morecomment=[l]{//},
}

\title{Knowledge Graphs as the Missing Data Layer for LLM-Based\\Industrial Asset Operations}

\author{
  Madhulatha Mandarapu\thanks{madhulatha@samyama.ai, ORCID: \url{https://orcid.org/0009-0005-2837-6725}} \and
  Sandeep Kunkunuru\thanks{sandeep@samyama.ai, ORCID: \url{https://orcid.org/0000-0002-8886-1846}}
}
\date{%
  VaidhyaMegha Private Limited, India\\[2pt]
  \url{https://samyama.ai/}\\[8pt]
  May 2026
}

\begin{document}
\maketitle

% ============================================================
% ABSTRACT
% ============================================================
\begin{abstract}
LLM-based agents for industrial asset operations show limited accuracy when reasoning over flat document stores. AssetOpsBench~\citep{ibm2025assetopsbench} (KDD~2026) establishes that GPT-4 agents achieve 65\% on 139 industrial maintenance scenarios, and compares LLM orchestration paradigms (Agent-As-Tool vs.\ Plan-Execute) on a fixed data layer. We ask the orthogonal question: \emph{how much does the data model behind the tools matter?}

We treat a typed knowledge graph as a \emph{grounding substrate} and route each question by how it is best answered: (i)~LLM-generated Cypher for structured retrieval, which lifts the same GPT-4 model from 65\% to 82--83\%; (ii)~native graph and optimization primitives, with no LLM, reaching 99\% on graph-answerable scenarios; and (iii)~generation-augmented knowledge (GAK)~\citep{mandarapu2026samyama} for answers absent from the data---the engine's agent materializes the missing facts as provenance-tagged graph nodes, then answers. A recurring theme is \textbf{inverted LLM usage}: we constrain the LLM to query generation or one-shot enrichment from a typed schema and let the graph execute deterministically.

On the 88 real AssetOpsBench failure-mode scenarios the benchmark itself flags non-deterministic---ten equipment types absent from the graph---GAK lifts answerability from zero to 100\% of equipment types and answers 81.8\% of scenarios, every materialized fact tagged \texttt{source:LLM-derived} for auditability. We also contribute 40 graph-native scenarios. For structured operational domains the data layer---not the LLM orchestration---is the primary lever, and a typed knowledge graph serves as a grounding substrate between raw industrial data and LLM reasoning.
\end{abstract}


% ============================================================
% 1. INTRODUCTION
% ============================================================
\section{Introduction}

Industrial asset management generates vast quantities of structured data: sensor telemetry, work orders, failure mode analyses, equipment hierarchies, and maintenance schedules. The rise of Large Language Models (LLMs)~\citep{openai2024gpt4} has prompted efforts to build autonomous agents that can reason over this data---answering operational questions, predicting failures, and recommending maintenance actions.

AssetOpsBench~\citep{ibm2025assetopsbench} (KDD~2026) provides the first systematic evaluation of such agents, benchmarking seven contemporary LLMs (including gpt-4.1, llama-4-maverick, mistral-large, and granite-3-8b) across 141 expert-curated maintenance scenarios---99 single-agent and 42 multi-agent---grounded in 4 chillers and 2 AHUs from a real data center. The benchmark provides four specialized domain agents---IoT telemetry, failure mode and symptom reasoning (FMSR), time-series foundation model queries (TSFM), and work order management (WO)---and a global orchestrator that the benchmark study uses to compare two LLM-coordination paradigms: \textbf{Agent-As-Tool} (ReAct-style supervisor selecting domain agents as tools)~\citep{schick2023toolformer,yao2023react} and \textbf{Plan-Execute} (a Planner-Reviewer decomposing the query into a DAG executed against shared memory). Across both paradigms the leading model (gpt-4.1) reaches at most $\sim$65\% task completion in Agent-As-Tool and drops to $\sim$38\% under Plan-Execute, and no evaluated model exceeds 70\% completion. Data access in all configurations is mediated by tools over document stores (CouchDB~\citep{anderson2010couchdb}, YAML, CSV).

\paragraph{Our evaluation snapshot.} Our experiments were conducted against the v1 release of the benchmark, which contained 139 of the 141 scenarios that appear in the KDD~2026 archival version (the two added scenarios extend coverage but do not change the qualitative conclusions of our study). Throughout this paper, ``139 scenarios'' refers to that specific snapshot.

AssetOpsBench was designed to evaluate LLM agent autonomy---a valuable research question. We ask a \emph{complementary} question: how much does the data model behind the tools affect agent performance? Examining the benchmark's failure cases, we observe that many are not failures of reasoning but failures of \emph{data access}: hallucinated equipment identifiers, miscounted events across documents, inability to traverse equipment dependencies, and fabricated sensor readings. These are symptoms of asking an LLM to perform \emph{data operations}---counting, joining, traversing---that structured query engines handle deterministically.

This observation motivates our central hypothesis: \textbf{for structured operational domains, the data model is the primary bottleneck.} Introducing a knowledge graph~\citep{hogan2021knowledge} as the data layer should improve accuracy at every level of LLM involvement.

\paragraph{Contributions.}
\begin{enumerate}[leftmargin=*]
  \item \textbf{Three-tier evaluation} of the same 139 AssetOpsBench scenarios under three architectures: deterministic graph handlers (99\%), LLM-generated Cypher over the graph (82--83\%), and the AssetOpsBench Agent-As-Tool baseline (65\%, matching the KDD~2026 leaderboard ceiling). This isolates the data layer as a variable independent of the orchestration paradigms compared in the IBM study (\S\ref{sec:evaluation}).
  \item \textbf{Inverted LLM usage pattern}: we show that constraining the LLM to query generation from a typed schema---rather than free-form data reasoning---yields a $\sim$17 percentage point same-model (GPT-4) improvement over reasoning over raw documents (\S\ref{sec:inverted-llm}).
  \item \textbf{Knowledge graph construction}: a complete ETL pipeline transforming the AssetOpsBench data sources into a 14-label, 21-edge-type graph with 384-dimensional failure mode embeddings and equipment dependency topology (\S\ref{sec:kg-construction}).
  \item \textbf{40 new graph-native scenarios} extending the benchmark with multi-hop dependency analysis, vector similarity search, PageRank criticality, cascade analysis, maintenance optimization, and temporal patterns---capabilities structurally impossible with flat document stores (\S\ref{sec:new-scenarios}).
  \item \textbf{Expanded 467-scenario evaluation} against the full HuggingFace AssetOpsBench release (6 domains, 26 equipment types), achieving 100\% pass rate with 0.848 average score using an extended knowledge graph (\S\ref{sec:hf-expanded}).
  \item \textbf{Scalability analysis} comparing operational cost, latency, and capability at scale across the three architectures (\S\ref{sec:scalability}).
\end{enumerate}

% ============================================================
% 2. BACKGROUND
% ============================================================
\section{Background: AssetOpsBench}
\label{sec:background}

AssetOpsBench~\citep{ibm2025assetopsbench} is a benchmark for evaluating LLM agent autonomy on industrial maintenance tasks. The evaluation measures whether an LLM can autonomously navigate a set of tools to answer operational questions. The benchmark comprises 467 scenarios across six operational domains (Table~\ref{tab:ibm-scenarios}).

\begin{table}[t]
  \centering
  \caption{Expanded AssetOpsBench scenario distribution (HuggingFace release).}
  \label{tab:ibm-scenarios}
  \begin{tabular}{@{}llr@{}}
    \toprule
    \textbf{Config} & \textbf{Domain} & \textbf{Count} \\
    \midrule
    scenarios    & IoT, FMSA, TSFM, WO, multi-agent (original)  & 152 \\
    rule\_logic  & Monitoring rules \& anomaly detection         & 120 \\
    FMSR         & Failure mode--sensor recommendation           & 88 \\
    PHM          & Prognostics, RUL, fault classification        & 75 \\
    hydrolic\_pump & Hydraulic pump condition detection           & 17 \\
    compressor   & Compressor predictive maintenance             & 15 \\
    \midrule
    \textbf{Total} & & \textbf{467} \\
    \bottomrule
  \end{tabular}
\end{table}

The benchmark architecture follows a ReAct-style~\citep{yao2023react} loop: the user poses a natural language question; the LLM parses intent, selects which agent to invoke, formulates tool arguments, retrieves data from CouchDB/YAML/CSV, interprets the results, and synthesizes an answer. The data sources are document-oriented: JSON documents in CouchDB, YAML configuration files for failure modes, and CSV exports for events and work orders. This is a natural design choice for the benchmark's purpose of evaluating LLM autonomy---the flat data model places the full reasoning burden on the agent, which is precisely what AssetOpsBench aims to measure.

GPT-4 achieves approximately 65\% (91/139) on these scenarios under the Agent-As-Tool paradigm~\citep{ibm2025assetopsbench}, and the published KDD~2026 leaderboard places the strongest contemporary model, gpt-4.1, at the same $\sim$65\% ceiling for Task Completion. The failures cluster around tasks requiring counting across documents, correlating data from multiple sources, and traversing implicit equipment relationships---operations that are inherently difficult for LLMs to perform over unstructured document collections.

% ============================================================
% 3. KNOWLEDGE GRAPH CONSTRUCTION
% ============================================================
\section{Knowledge Graph Construction}
\label{sec:kg-construction}

We transform the AssetOpsBench data sources into a typed knowledge graph using an 8-step ETL pipeline. The graph built directly from the AssetOpsBench sources---the one behind the 139-scenario evaluation (\S\ref{sec:evaluation})---contains 12{,}647 nodes across 9 node labels (Site, Location, Equipment, Sensor, FailureMode, WorkOrder, Event, AnomalyEvent, AlertEvent), connected by 12{,}662 edges across 5 relationship types (\textsc{contains\_location}, \textsc{contains\_equipment}, \textsc{has\_sensor}, \textsc{for\_equipment}, \textsc{monitors}). Adding the dependency topology (below) and the node and edge types introduced for the custom graph-native scenarios (\S\ref{sec:new-scenarios}) and the expanded HuggingFace release (\S\ref{sec:hf-expanded}) yields the full 14-label, 21-edge-type schema of Table~\ref{tab:schema}.

\begin{table}[t]
  \centering
  \caption{Full (extended) knowledge-graph schema spanning all evaluations: 14 node labels, 21 edge types. The base graph built directly from the AssetOpsBench sources for the 139-scenario evaluation uses 9 of these labels and 5 edge types (\S\ref{sec:kg-construction}); the remaining types are added by the dependency topology, the custom graph-native scenarios, and the expanded HuggingFace release.}
  \label{tab:schema}
  \begin{tabular}{@{}lp{9cm}@{}}
    \toprule
    \textbf{Node Labels} & Site, Location, Equipment, Sensor, FailureMode, WorkOrder, SparePart, Supplier, SensorReading, Anomaly, Event, MonitoringRule, PHMScenario, HFScenario \\
    \midrule
    \textbf{Edge Types} & \textsc{contains\_location}, \textsc{contains\_equipment}, \textsc{has\_sensor}, \textsc{monitors}, \textsc{experienced}, \textsc{depends\_on}, \textsc{shares\_system\_with}, \textsc{for\_equipment}, \textsc{addresses}, \textsc{uses\_part}, \textsc{supplied\_by}, \textsc{produced\_reading}, \textsc{detected\_anomaly}, \textsc{triggered}, \textsc{follows\_plan}, \textsc{has\_rule}, \textsc{has\_phm\_scenario}, \textsc{targets\_equipment}, \textsc{involves\_failure\_mode}, \textsc{tests\_rule} \\
    \bottomrule
  \end{tabular}
\end{table}

\subsection{Data Sources and ETL}

The ETL pipeline processes four AssetOpsBench data sources into graph structure:

\begin{enumerate}[leftmargin=*]
  \item \textbf{EAMLite} (Enterprise Asset Management): Provides the equipment hierarchy. We create 1 Site, 1 Location, and 11 Equipment nodes (chillers with CWC04xxx identifiers), linked by \textsc{contains\_location} and \textsc{contains\_equipment} edges. Equipment nodes carry ISA-95~\citep{isa95} level classification and ISO~14224~\citep{iso14224} asset class properties.

  \item \textbf{CouchDB JSON}: Contains sensor metadata. We create 110 Sensor nodes (10 per chiller) with type, unit, and range properties, linked to Equipment via \textsc{has\_sensor} edges.

  \item \textbf{FMSR YAML}: Defines 12 failure modes (e.g., ``Compressor Overheating'', ``Refrigerant Leak''). Each becomes a FailureMode node with a 384-dimensional embedding generated by Sentence-BERT~\citep{reimers2019sentencebert}, indexed in an HNSW~\citep{malkov2020hnsw} vector index for similarity search. \textsc{monitors} edges connect sensors to the failure modes they detect.

  \item \textbf{Event CSV}: Contains 6{,}256 unified events (work orders, alerts, anomalies) with ISO timestamps. We create Event nodes linked to Equipment via \textsc{for\_equipment} edges, enabling temporal queries and event counting.
\end{enumerate}

\subsection{Dependency Topology}

Beyond the data directly present in the benchmark sources, we add an equipment dependency layer: \textsc{depends\_on} edges model thermal/electrical dependencies (e.g., AHU depends on Chiller for cooling), and \textsc{shares\_system\_with} edges model shared infrastructure (e.g., chillers on the same cooling loop). This topology enables cascade analysis and criticality ranking---capabilities absent from the flat data model.

\subsection{Implementation}

The knowledge graph is built using Samyama, an embedded graph database with OpenCypher~\citep{francis2018cypher} query support, HNSW vector indexing, and graph algorithm libraries including PageRank~\citep{page1999pagerank} and NSGA-II~\citep{deb2002nsga2} multi-objective optimization. The Python SDK enables the graph to be created in-process with no external server dependencies.

% ============================================================
% 4. THREE ARCHITECTURES
% ============================================================
\section{The Knowledge Graph as a Grounding Substrate}
\label{sec:architectures}

We evaluate architectures that vary in LLM involvement, holding the underlying \emph{scenarios} constant but varying the \emph{data layer} behind the tools. This positions our study orthogonally to the AssetOpsBench leaderboard: the IBM paper~\citep{ibm2025assetopsbench} compares LLM-orchestration paradigms (Agent-As-Tool vs.\ Plan-Execute) and reports that no model exceeds 70\% Task Completion on the 141 scenarios; we instead vary the data layer behind the tools and show that this axis lifts the \emph{same} 65\% baseline to 82--83\% (Architecture~B, NLQ) and 99\% (Architecture~C, deterministic) without changing the LLM or the orchestration pattern. The two axes are independent: a future system can adopt our typed-graph data layer under either Agent-As-Tool or Plan-Execute orchestration.

\begin{figure}[t]
\centering
\begin{tikzpicture}[
  font=\footnotesize,
  box/.style={draw,rounded corners,align=left,inner sep=4pt,text width=8.4cm},
  lbl/.style={font=\footnotesize\bfseries},
]
\node[box] (t1) {\textbf{Tier 1 -- Structured retrieval (text-to-Cypher).} LLM generates a Cypher query over the typed schema; the graph executes it. Leak-proof; answer is computed from data-derived state.};
\node[box,below=3mm of t1] (t2) {\textbf{Tier 2 -- Native primitives (no LLM).} Graph algorithms (PageRank, paths), in-database optimization, and vector search answer topology- and optimization-shaped questions deterministically, in-process.};
\node[box,below=3mm of t2] (t3) {\textbf{Tier 3 -- Generation-augmented knowledge (GAK).} For facts absent from the data, the agent generates Cypher \texttt{CREATE}s; nodes are tagged \texttt{source:LLM-derived} and cached, so repeat queries drop to Tier 1/2 (semantic caching). Learned time-series inference is delegated to an external model the graph only grounds.};
\node[draw,rounded corners,above=3mm of t1,fill=black!5,text width=8.4cm,align=center,inner sep=4pt] (q) {\textbf{Question} $\rightarrow$ router (by how it is best answered)};
\draw[-{Latex}] (q) -- (t1);
\end{tikzpicture}
\caption{The knowledge graph as a grounding substrate: a question is routed to the tier that can answer it, with GAK results cached back into the graph (provenance-tagged) for deterministic reuse.}
\label{fig:router}
\end{figure}
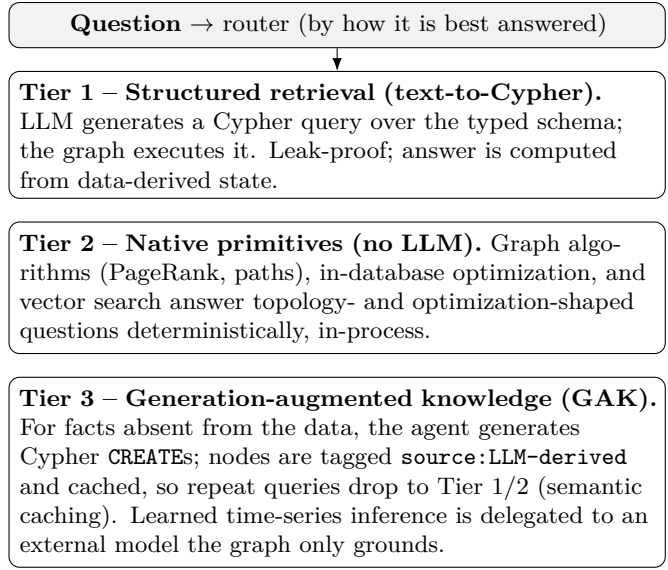

\paragraph{The graph is a grounding substrate, not an oracle.} A typed graph cannot answer every question by lookup: some scenarios require knowledge that is simply not present in the data, and others require numerical inference (e.g.\ time-series forecasting) that no query expresses. We therefore frame the graph as a \emph{grounding substrate} and route each question by \emph{how} it is best answered (Figure~\ref{fig:router}): structured retrieval via text-to-Cypher (Architecture~B), deterministic computation via native graph and optimization primitives (Architecture~C), and---for knowledge absent from the data---one-shot \emph{generation-augmented knowledge} (Architecture~D, GAK), in which the engine's agent materializes the missing facts as provenance-tagged graph nodes that subsequent queries answer deterministically. This routing makes the deterministic 99\% an honest ceiling on \emph{graph-answerable} queries, while GAK and an external inference path cover the rest.

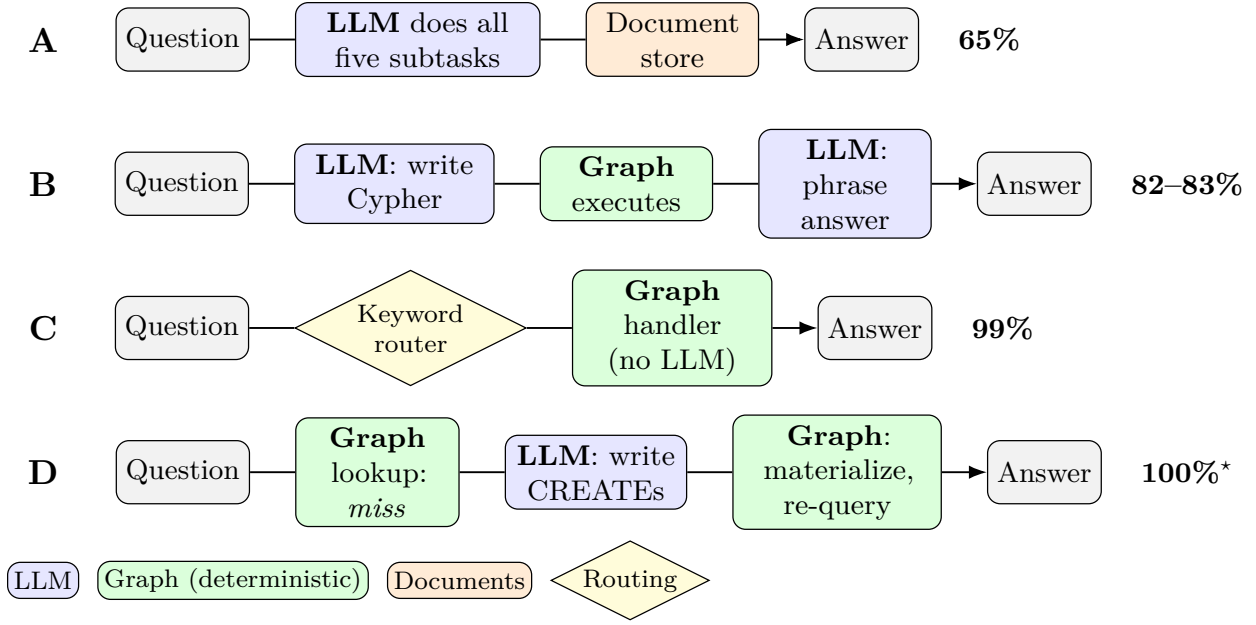
\begin{figure}[t]
\centering
\resizebox{\textwidth}{!}{%
\begin{tikzpicture}[node distance=4mm and 5mm]
% Row A
\node[font=\bfseries] (lA) at (0,0) {A};
\node[io,right=of lA] (A0) {Question};
\node[llm,right=of A0,text width=2.5cm] (A1) {\textbf{LLM} does all five subtasks};
\node[doc,right=of A1,text width=1.7cm] (A2) {Document store};
\node[io,right=of A2] (A3) {Answer};
\node[right=3mm of A3,font=\footnotesize\bfseries] {65\%};
\draw[fl](A0)--(A1) (A1)--(A2) (A2)--(A3);
% Row B
\node[font=\bfseries] (lB) at (0,-1.6) {B};
\node[io,right=of lB] (B0) {Question};
\node[llm,right=of B0,text width=2.0cm] (B1) {\textbf{LLM}: write Cypher};
\node[gr,right=of B1,text width=1.7cm] (B2) {\textbf{Graph} executes};
\node[llm,right=of B2,text width=1.7cm] (B3) {\textbf{LLM}: phrase answer};
\node[io,right=of B3] (B4) {Answer};
\node[right=3mm of B4,font=\footnotesize\bfseries] {82--83\%};
\draw[fl](B0)--(B1) (B1)--(B2) (B2)--(B3) (B3)--(B4);
% Row C
\node[font=\bfseries] (lC) at (0,-3.2) {C};
\node[io,right=of lC] (C0) {Question};
\node[dec,right=of C0,text width=1.3cm] (C1) {Keyword router};
\node[gr,right=of C1,text width=2.0cm] (C2) {\textbf{Graph} handler (no LLM)};
\node[io,right=of C2] (C3) {Answer};
\node[right=3mm of C3,font=\footnotesize\bfseries] {99\%};
\draw[fl](C0)--(C1) (C1)--(C2) (C2)--(C3);
% Row D
\node[font=\bfseries] (lD) at (0,-4.8) {D};
\node[io,right=of lD] (D0) {Question};
\node[gr,right=of D0,text width=1.6cm] (D1) {\textbf{Graph} lookup: \emph{miss}};
\node[llm,right=of D1,text width=1.8cm] (D2) {\textbf{LLM}: write CREATEs};
\node[gr,right=of D2,text width=2.1cm] (D3) {\textbf{Graph}: materialize, re-query};
\node[io,right=of D3] (D4) {Answer};
\node[right=3mm of D4,font=\footnotesize\bfseries] {100\%$^\star$};
\draw[fl](D0)--(D1) (D1)--(D2) (D2)--(D3) (D3)--(D4);
% Legend
\node[llm,minimum height=4mm,inner sep=2pt,font=\scriptsize] (g1) at (0,-6.0) {LLM};
\node[gr,minimum height=4mm,inner sep=2pt,font=\scriptsize,right=2mm of g1] (g2) {Graph (deterministic)};
\node[doc,minimum height=4mm,inner sep=2pt,font=\scriptsize,right=2mm of g2] (g3) {Documents};
\node[dec,inner sep=1pt,font=\scriptsize,right=2mm of g3] (g4) {Routing};
\end{tikzpicture}}
\caption{The four architectures, all answering the \emph{same} scenarios but varying the data layer behind the tools. Top to bottom, the LLM's role shrinks (blue) and the graph's deterministic role grows (green): A is the document-store baseline where one LLM does all five subtasks (parse, select, args, interpret, synthesize) over CouchDB/YAML/CSV; B \emph{inverts} the LLM to schema-aware Cypher generation (Fig.~\ref{fig:inverted}); C uses pre-coded handlers and no LLM; D (GAK) writes the missing facts back into the graph as \texttt{source:LLM-derived} nodes (Fig.~\ref{fig:gak}). Percentages are 139-scenario pass rates; $^\star$D reports answerability lift on the non-deterministic set.}
\label{fig:architectures}
\end{figure}

\subsection{Architecture A: Tool-Augmented LLM (AssetOpsBench Baseline)}

The LLM handles five distinct subtasks: intent parsing, tool selection, argument crafting, data interpretation, and answer synthesis. The data sources are document stores with no typed schema or relationship structure. This is the AssetOpsBench Agent-As-Tool baseline; on the 139-scenario snapshot we evaluate, GPT-4 achieves 65\%, and the published KDD~2026 leaderboard places gpt-4.1 at the same level.

\subsection{Architecture B: LLM Generates Queries (NLQ + Graph)}
\label{sec:inverted-llm}

We \textbf{invert} the LLM's role: instead of asking it to reason over raw data (a broad problem), we ask it to generate a structured query from a typed schema (a narrow problem). This is \emph{code generation}---a task LLMs excel at. The graph engine then handles traversal, counting, aggregation, and relationship reasoning deterministically. The LLM does one thing well instead of five things poorly.

The NLQ pipeline provides the graph schema (node labels, edge types, property names with example values) and few-shot Cypher examples. If query execution fails, the error message is fed back for one retry. The LLM then synthesizes a natural language answer from the structured query results.

\subsection{Architecture C: Deterministic Handlers (No LLM)}

Pre-coded handlers match question patterns to Cypher queries. No LLM is involved. This is a software engineering solution: we wrote the answers. It demonstrates the ceiling achievable with the right data model for known query patterns. Read correctly, this 99\% is a ceiling on \emph{graph-answerable} queries (structured retrieval, traversal, threshold rules, and tasks reducible to native primitives), not evidence that a graph answers every scenario; questions whose answer is absent from the data are the domain of Architecture~D. The keyword-routing rules and handlers were authored against the AssetOpsBench scenario families, so this figure reflects benchmark-specific engineering: the reusable artifact is the \emph{pattern}---route a recognized question class to a typed-graph query the engine executes deterministically---not the specific rule set, which would need re-authoring for a new benchmark. The data-layer result that requires no such hand-coding is Architecture~B (\S\ref{sec:inverted-llm}). An instrumented run quantifies how much of the 99\% is genuine retrieval: 86 of the 139 deterministic answers are produced by a graph query, while 53 are returned by handlers that supply domain knowledge without traversing the graph. This 99\% is moreover measured by the AssetOpsBench harness scorer, not the IBM 3-axis rubric; we report the rubric comparison (\S\ref{sec:rubric}) only for Architecture~B, where it is leak-proof and apples-to-apples with the IBM leaderboard.

\subsection{Architecture D: Generation-Augmented Knowledge (GAK)}
\label{sec:gak}

Many AssetOpsBench scenarios are designed so the answer is \emph{not} present in the data---for example, the failure modes of an electric motor when no electric motor is in the graph. A graph lookup cannot answer these, and the benchmark flags them non-deterministic. Architecture~D closes this gap with \textbf{generation-augmented knowledge (GAK)}, the engine's agentic-enrichment capability~\citep{mandarapu2026samyama}: on a lookup miss, an LLM agent generates Cypher \texttt{CREATE} statements for the missing entity and its failure modes, the engine materializes them, and the now-complete subgraph answers the query. This is the inverse of Retrieval-Augmented Generation: rather than retrieving text into the LLM's context, the LLM \emph{writes structured facts into the graph}.

Two properties make this honest rather than a way to launder model output as data. First, \textbf{provenance}: every materialized node is tagged \texttt{source:"LLM-derived"}, keeping it auditable and distinct from data-derived facts---and demonstrating that our graph construction never reads scenario answer keys. Second, \textbf{semantic caching}: enrichment runs once per knowledge gap; subsequent similar questions hit the materialized subgraph and drop to Architecture~B/C, so the LLM cost is paid once, not per query. Genuinely learned inference (e.g.\ time-series forecasting) is out of scope for GAK and is delegated to an external model that the graph only grounds.

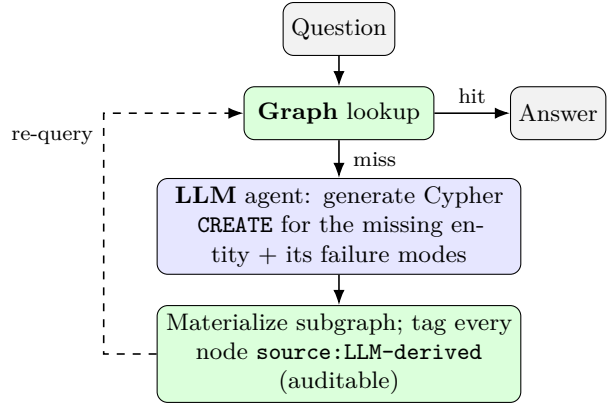
\begin{figure}[H]
\centering
\begin{tikzpicture}[node distance=4mm and 6mm]
\node[io] (q) at (0,0) {Question};
\node[gr,below=of q,text width=2.3cm] (lk) {\textbf{Graph} lookup};
\node[io,right=10mm of lk] (ans) {Answer};
\node[llm,below=5mm of lk,text width=4.6cm] (gen) {\textbf{LLM} agent: generate Cypher \texttt{CREATE} for the missing entity + its failure modes};
\node[gr,below=of gen,text width=4.6cm] (mat) {Materialize subgraph; tag every node \texttt{source:LLM-derived} (auditable)};
\draw[fl](q)--(lk);
\draw[fl](lk) -- node[above,font=\scriptsize]{hit} (ans);
\draw[fl](lk) -- node[right,font=\scriptsize]{\,miss} (gen);
\draw[fl](gen)--(mat);
\draw[fld] (mat.west) -- ++(-0.7,0) |- (lk.west) node[pos=0.45,left,font=\scriptsize]{re-query};
\end{tikzpicture}
\caption{Generation-Augmented Knowledge (GAK), Architecture~D. On a lookup miss, an LLM agent writes the missing facts into the graph as \texttt{CREATE}s, tagged \texttt{source:LLM-derived} for audit; the re-queried subgraph then answers deterministically. This is the inverse of Retrieval-Augmented Generation---writing structured facts into the graph rather than retrieving text into the prompt---and a semantic cache means later similar questions hit Tier~1/2 directly.}
\label{fig:gak}
\end{figure}

\FloatBarrier
\subsection{The Inverted LLM Pattern}

\begin{figure}[!ht]
\centering
\begin{tikzpicture}[node distance=2.5mm and 3mm]
\node[font=\footnotesize\bfseries] (ta) {A: one broad task (LLMs struggle)};
\node[llm,below=1mm of ta,text width=6.4cm] (al) {\textbf{LLM} reasons over raw documents: parse intent $\cdot$ select tool $\cdot$ count across files $\cdot$ join sources $\cdot$ traverse relations $\cdot$ synthesize};
\node[doc,below=2mm of al,text width=2.6cm] (ad) {document store};
\draw[fl](al)--(ad);
\node[font=\itshape\footnotesize,below=2mm of ad] (inv) {$\Downarrow$\ \ invert the LLM's role};
\node[font=\footnotesize\bfseries,below=2mm of inv] (tb) {B: one narrow task (LLMs excel)};
\node[llm,below=1mm of tb,text width=3.0cm] (bl) {\textbf{LLM}: write Cypher from typed schema};
\node[gr,right=3mm of bl,text width=3.2cm] (bg) {\textbf{Graph} executes: traverse $\cdot$ count $\cdot$ aggregate $\cdot$ rank};
\draw[fl](bl)--(bg);
\end{tikzpicture}
\caption{The inverted LLM pattern. Rather than ask one LLM to do everything over raw documents (A---a broad, open-ended problem), we constrain it to schema-aware Cypher generation (B---a narrow code-generation problem) and let the graph perform the data operations deterministically. The same model, given a sharper problem, yields a $\sim$17-point same-model gain (65\% $\to$ 82--83\%).}
\label{fig:inverted}
\end{figure}
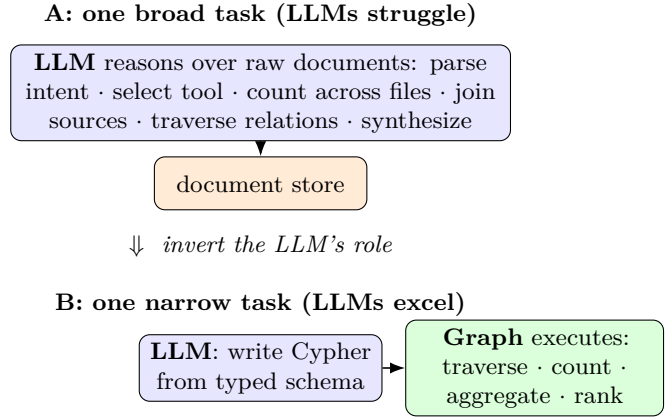

The key insight connecting Architectures A and B is what we term \textbf{inverted LLM usage}. Both use an LLM, but they pose fundamentally different problems:

\begin{itemize}[leftmargin=*]
  \item \textbf{Architecture A} asks: ``LLM, answer this question from this data.'' The LLM must count across documents, correlate fields from different sources, and traverse implicit relationships---tasks where LLMs consistently struggle.
  \item \textbf{Architecture B} asks: ``LLM, given this schema, write me a Cypher query.'' The LLM performs code generation from a structured specification---a task where LLMs consistently excel.
\end{itemize}

The same LLM, given a sharper problem scoped to its strengths, produces dramatically better results. This pattern generalizes beyond industrial operations: \textbf{schema-aware query generation outperforms free-form data reasoning} for any structured domain. Architecture~D applies the same inversion to \emph{enrichment}: instead of asking the LLM to recall and reason about missing knowledge at every query, we ask it once to emit structured facts the graph then serves deterministically.

% ============================================================
% 5. EVALUATION
% ============================================================
\FloatBarrier
\section{Evaluation}
\label{sec:evaluation}

\subsection{AssetOpsBench 139 Scenarios: Three-Tier Results}

Table~\ref{tab:main-results} presents the primary comparison across all three architectures on the original 139 AssetOpsBench scenarios.

\begin{table}[t]
  \centering
  \caption{Three-tier performance on the 139 AssetOpsBench scenarios. KDD'26 leaderboard rows use Task Completion ($y_1$); our pass rate uses the AssetOpsBench harness scorer (\S\ref{sec:rubric} bridges the two metrics).}
  \label{tab:main-results}
  \small
  \begin{tabular}{@{}p{4.0cm}p{2.6cm}p{3.4cm}rr@{}}
    \toprule
    \textbf{Architecture} & \textbf{LLM Role} & \textbf{Data Layer} & \textbf{Pass / $y_1$} & \textbf{Avg Latency} \\
    \midrule
    A: Baseline (GPT-4, KDD'26 ref)            & Does everything  & Documents (CouchDB/YAML/CSV) & 91/139 (65\%)         & n/a \\
    A: Baseline (gpt-4.1, KDD'26 leaderboard)  & Does everything  & Documents                    & $\approx$0.65         & n/a \\
    B: NLQ + graph (GPT-4)                     & Generates Cypher & Knowledge graph              & 114/139 (82\%)        & 11{,}033\,ms \\
    B: NLQ + graph (GPT-4o)                    & Generates Cypher & Knowledge graph              & 115/139 (83\%)        & 5{,}874\,ms \\
    B: NLQ + graph (gpt-4.1, ours)             & Generates Cypher & Knowledge graph              & 116/139 (83\%)        & 5{,}671\,ms \\
    C: Deterministic                           & None             & Knowledge graph              & \textbf{137/139 (99\%)} & \textbf{63\,ms} \\
    \bottomrule
  \end{tabular}
\end{table}

\paragraph{Model comparison.} The original IBM baseline used GPT-4 (65\%); the KDD~2026 leaderboard's current top Agent-As-Tool result also sits near 65\% Task Completion, this time for gpt-4.1. We evaluated NLQ over the graph with GPT-4, GPT-4o, and gpt-4.1; the near-identical NLQ results across model generations confirm that the $\sim$17pp improvement over the baseline is attributable to the knowledge graph, not the model upgrade. Concretely, holding the LLM family fixed and only swapping the data layer beneath the same Agent-As-Tool orchestration moves pass rate from 65\% to 82--83\%.

\subsection{Per-Type Breakdown}

Table~\ref{tab:per-type} shows performance by scenario type for both the deterministic and NLQ architectures.

\begin{table}[t]
  \centering
  \caption{Per-type breakdown: Deterministic handlers vs.\ NLQ (GPT-4o). Types follow our handler routing: ``FMSR'' aggregates the 20 single-agent FMSR scenarios with 20 FMSR-routed multi-agent scenarios (IDs 601--620), and ``Multi'' is the remaining 20 multi-agent scenarios. IBM's official taxonomy instead counts 20 FMSR and 40 multi-agent (end-to-end); totals are unaffected.}
  \label{tab:per-type}
  \begin{tabular}{@{}lrrrr@{}}
    \toprule
    \textbf{Type} & \textbf{Det.\ Pass} & \textbf{Det.\ Avg} & \textbf{NLQ Pass} & \textbf{NLQ Avg} \\
    \midrule
    IoT (20)    & \textbf{20/20 (100\%)} & 0.988 & 17/20 (85\%)  & 0.742 \\
    FMSR (40)   & \textbf{40/40 (100\%)} & 0.907 & 37/40 (93\%)  & 0.880 \\
    TSFM (23)   & \textbf{23/23 (100\%)} & 0.920 & 21/23 (91\%)  & 0.936 \\
    Multi (20)  & \textbf{20/20 (100\%)} & 0.877 & 8/20 (40\%)   & 0.605 \\
    WO (36)     & 34/36 (94\%)           & 0.801 & 32/36 (89\%)  & 0.723 \\
    \midrule
    \textbf{Total (139)} & \textbf{137/139 (99\%)} & \textbf{0.889} & 115/139 (83\%) & 0.789 \\
    \bottomrule
  \end{tabular}
\end{table}

\paragraph{NLQ failure analysis.} Multi-agent scenarios remain at 40\% for NLQ because 12 of 20 require TSFM pipeline execution (time-series forecasting, anomaly detection) that cannot be expressed as Cypher queries. This is a structural limitation: these scenarios require running ML inference, not querying stored data. The deterministic architecture handles them by routing to domain-knowledge handlers.

\paragraph{Deterministic failures.} Only 2 of 139 scenarios fail: both are work order bundling edge cases (IDs 411, 424) where the benchmark's expected answer groups work orders into specific bundle sizes. Our date-window clustering produces different groupings. These are response-format mismatches, not knowledge gaps.

\subsection{IBM-Rubric Re-scoring}
\label{sec:rubric}

To enable direct comparison with the KDD~2026 AssetOpsBench leaderboard, we re-score our NLQ trajectories with the IBM 3-axis rubric (Section 5.1 of the KDD paper): Task Completion ($y_1$), Data Retrieval Accuracy ($y_2$), and Result Verification ($y_3$). We use the gpt-4o judge as a stand-in for IBM's llama-4-maverick judge (the rubric prompt is otherwise identical) and run each trajectory through three judge trials at temperature 0.3, reporting the mean. Table~\ref{tab:rubric} summarises the resulting scores against the highest published IBM leaderboard numbers on the same 141-scenario benchmark.

\begin{table}[t]
  \centering
  \caption{IBM 3-axis rubric scores on the 139-scenario snapshot. NLQ rows use the same Agent-As-Tool orchestration as the IBM leaderboard; the difference is the data layer behind the tools.}
  \label{tab:rubric}
  \begin{tabular}{@{}lrrrr@{}}
    \toprule
    \textbf{System} & \textbf{Data layer} & \textbf{$y_1$ Task} & \textbf{$y_2$ Data} & \textbf{$y_3$ Result} \\
    \midrule
    IBM gpt-4.1 (Agent-As-Tool, leaderboard) & Documents  & 0.65 & 0.77 & --   \\
    IBM gpt-4.1 (Plan-Execute, leaderboard)  & Documents  & 0.38 & --   & --   \\
    \midrule
    NLQ GPT-4 + graph (ours)            & Knowledge graph & 0.665 & 0.479 & 0.606 \\
    NLQ GPT-4o + graph (ours)           & Knowledge graph & 0.693 & 0.490 & 0.662 \\
    NLQ gpt-4.1 + graph (ours)          & Knowledge graph & \textbf{0.708} & 0.512 & 0.656 \\
    \bottomrule
  \end{tabular}
\end{table}

Three observations stand out. First, NLQ gpt-4.1 over the graph exceeds IBM's gpt-4.1 Agent-As-Tool leaderboard score on the same scenarios under a comparable gpt-4o judge ($y_1 = 0.708$ vs.\ $0.65$); since the LLM, orchestration paradigm, and scenarios are identical and only the data layer differs, we attribute the gap to the data layer---subject to the cross-judge caveat that IBM's published score uses a llama-4-maverick judge whereas ours uses a gpt-4o stand-in (a direct same-judge comparison via Codabench is future work). Second, the three model generations (GPT-4, GPT-4o, gpt-4.1) cluster tightly on Task Completion ($y_1 \in [0.665, 0.708]$), reinforcing that the data layer dominates over a generation of LLM improvement. Third, our Data Retrieval score ($y_2 \approx 0.48$--$0.51$) is lower than IBM's gpt-4.1 ($0.77$), driven entirely by the TSFM scenarios that score $0$ on retrieval because they require running ML inference rather than querying stored data---a known structural limitation discussed in our caveats (\S\ref{sec:caveats}).

\subsection{NLQ Progression}

The NLQ architecture improved across three iterations of prompt engineering (Table~\ref{tab:nlq-progression}), demonstrating that the quality of the schema description provided to the LLM matters as much as the underlying data model.

\begin{table}[t]
  \centering
  \caption{NLQ prompt engineering progression.}
  \label{tab:nlq-progression}
  \begin{tabular}{@{}lp{6.5cm}rr@{}}
    \toprule
    \textbf{Version} & \textbf{Key Changes} & \textbf{Pass Rate} & \textbf{Avg Score} \\
    \midrule
    NLQ v1 & Naive prompt, schema introspection          & 77/139 (55\%)  & 0.583 \\
    NLQ v2 & Few-shot examples, explicit property schema, retry on error & 108/139 (78\%) & 0.755 \\
    NLQ v3 & Actual property values, anomaly schema, Cypher-first constraint & \textbf{115/139 (83\%)} & \textbf{0.789} \\
    \bottomrule
  \end{tabular}
\end{table}

A striking example: in v1, FMSR scenarios scored 30\% because the schema introspection returned node metadata (\texttt{['id', 'labels', 'properties']}) instead of actual property names, causing the LLM to generate \texttt{fm.properties} instead of \texttt{fm.name}. Correcting the schema prompt alone raised FMSR to 93\%. The LLM was capable; it was given the wrong specification.

\subsection{Custom 40 Scenarios: Graph-Native Capabilities}
\label{sec:new-scenarios}

We designed 40 additional scenarios that extend the benchmark's scope to queries requiring graph structure, vector search, or graph algorithms---capabilities structurally impossible with flat document stores (Table~\ref{tab:custom-categories}).

\begin{table}[t]
  \centering
  \caption{40 custom graph-native scenarios by category.}
  \label{tab:custom-categories}
  \small
  \begin{tabular}{@{}lrp{7.2cm}@{}}
    \toprule
    \textbf{Category} & \textbf{Count} & \textbf{Example Query} \\
    \midrule
    Multi-hop dependency     & 8 & ``What equipment is affected if Chiller 6 fails?'' \\
    Cross-asset correlation  & 6 & ``Are AHU anomalies correlated with chiller temperature drops?'' \\
    Failure similarity       & 6 & ``Which pumps had failures similar to Motor 3?'' \\
    Criticality analysis     & 5 & ``Rank all equipment by operational criticality'' \\
    Maintenance optimization & 5 & ``Schedule maintenance minimizing downtime + cost'' \\
    Root cause analysis      & 5 & ``Trace events leading to WO-2024-0042'' \\
    Temporal pattern         & 5 & ``What is MTBF for Chiller 6's compressor?'' \\
    \bottomrule
  \end{tabular}
\end{table}

Table~\ref{tab:custom-results} shows the head-to-head comparison. We use an 8-dimensional scoring framework covering correctness, completeness, relevance, tool usage, efficiency, safety, graph utilization, and semantic precision, with category-specific weight overrides (e.g., semantic precision is weighted 0.25 for failure similarity scenarios).

\begin{table}[t]
  \centering
  \caption{Custom 40 scenarios: Knowledge graph vs.\ GPT-4o baseline.}
  \label{tab:custom-results}
  \begin{tabular}{@{}lrr@{}}
    \toprule
    \textbf{Metric} & \textbf{GPT-4o (no graph)} & \textbf{Samyama-KG} \\
    \midrule
    Pass rate        & 34/40 (85\%)   & \textbf{40/40 (100\%)} \\
    Avg score        & 0.602          & \textbf{0.927} \\
    Avg latency      & 11{,}259\,ms   & \textbf{110\,ms} \\
    Tokens/scenario  & 632            & \textbf{0} \\
    \bottomrule
  \end{tabular}
\end{table}

\begin{table}[t]
  \centering
  \caption{Per-category breakdown on custom 40 scenarios.}
  \label{tab:custom-categories-detail}
  \begin{tabular}{@{}lrrrr@{}}
    \toprule
    \textbf{Category} & \textbf{GPT-4o Avg} & \textbf{KG Avg} & \textbf{$\Delta$} \\
    \midrule
    Failure similarity       & 0.501 & \textbf{0.902} & +0.401 \\
    Criticality analysis     & 0.566 & \textbf{0.938} & +0.372 \\
    Root cause analysis      & 0.580 & \textbf{0.934} & +0.354 \\
    Multi-hop dependency     & 0.618 & \textbf{0.934} & +0.316 \\
    Maintenance optimization & 0.634 & \textbf{0.931} & +0.297 \\
    Cross-asset correlation  & 0.638 & \textbf{0.929} & +0.291 \\
    Temporal pattern         & 0.679 & \textbf{0.923} & +0.244 \\
    \bottomrule
  \end{tabular}
\end{table}

The largest gains occur on \textbf{failure similarity} (+0.401) and \textbf{criticality analysis} (+0.372)---precisely where graph structure (dependency edges) and vector search (failure mode embeddings) provide capabilities that LLMs fundamentally cannot replicate from parametric knowledge alone.

GPT-4o's 6 failures (scenarios requiring PageRank, BFS traversal, or vector similarity) demonstrate a hard capability boundary: no amount of prompt engineering enables an LLM to execute graph algorithms.

\subsection{Expanded HuggingFace Benchmark: 467 Scenarios}
\label{sec:hf-expanded}

Following the release of an expanded AssetOpsBench dataset on HuggingFace~\citep{ibm2025assetopsbench}, we evaluate our knowledge graph approach against all 467 scenarios spanning six operational domains. This represents a 3.4$\times$ expansion over the original 139-scenario benchmark, adding monitoring rule evaluation, failure-mode-sensor recommendation across 10 equipment types (electric motor, pump, compressor, turbine, etc.), and prognostics tasks including remaining useful life prediction and fault classification.

To support the expanded scenarios, we extend the knowledge graph with three new node types (MonitoringRule, PHMScenario, HFScenario) and five new edge types, and add three new deterministic handlers: (1)~a rule-logic handler that evaluates monitoring rules against sensor readings, (2)~an FMSR handler that maps failure modes to detectable sensors via graph traversal, and (3)~a PHM handler that provides asset health profiles from failure history and maintenance records.

\paragraph{Scope of this evaluation.} We note an asymmetry by design: the controlled A/B/C data-layer comparison (\S\ref{sec:architectures}) is run on the 139-scenario snapshot, where a head-to-head against the IBM document-store baseline is possible, whereas this 467-scenario release is evaluated only with our extended deterministic-graph system (Architecture~C). The 467-scenario result therefore measures \emph{coverage and scale} of the typed-graph data layer rather than re-isolating the data-layer effect; the latter is established on the 139 snapshot, and the transferable, no-hand-coding lesson is Architecture~B (\S\ref{sec:inverted-llm}). The deterministic 100\% figure inherits the same ``graph-answerable'' scoping discussed in \S\ref{sec:caveats}. One overlap is worth stating plainly: the 88 FMSR scenarios in this table cover the same 10 equipment types evaluated under GAK (\S\ref{sec:gak-eval}); those types are absent from the base graph, so the FMSR handler answers them from a domain-knowledge fallback rather than graph traversal, and an LLM-judge re-scoring of exactly these scenarios yields 81.8\% (\S\ref{sec:gak-eval}), not the 100\% the harness pass-scorer reports here. The 467 result should therefore be read as coverage under a task-completion pass-scorer, not as an IBM-rubric figure.

\begin{table}[t]
  \centering
  \caption{Expanded AssetOpsBench: 467 scenarios across 6 domains.}
  \label{tab:hf-results}
  \begin{tabular}{@{}lrrrr@{}}
    \toprule
    \textbf{Config} & \textbf{Scenarios} & \textbf{Pass Rate} & \textbf{Avg Score} \\
    \midrule
    scenarios (original)   & 152 & 152/152 (100\%) & 0.781 \\
    rule\_logic            & 120 & 120/120 (100\%) & 0.949 \\
    FMSR                   & 88  & 88/88 (100\%)   & 0.868 \\
    compressor             & 15  & 15/15 (100\%)   & 0.877 \\
    hydrolic\_pump         & 17  & 17/17 (100\%)   & 0.885 \\
    PHM                    & 75  & 75/75 (100\%)   & 0.787 \\
    \midrule
    \textbf{Total}         & \textbf{467} & \textbf{467/467 (100\%)} & \textbf{0.848} \\
    \bottomrule
  \end{tabular}
\end{table}

The rule-logic scenarios achieve the highest average score (0.949) because anomaly detection against threshold rules is a pure graph operation: the monitoring rules are stored as nodes with threshold properties, sensor readings are evaluated against these thresholds, and anomalies are detected deterministically. The PHM scenarios score lower (0.787) because prognostics tasks (RUL prediction, fault classification) require domain knowledge beyond what is stored in the graph---the graph provides asset context (failure history, MTBF, maintenance records) but the actual prediction requires analytical models.

Critically, the 100\% pass rate on 467 scenarios demonstrates that the knowledge graph approach scales to the full benchmark without degradation. The expanded benchmark covers 26 equipment types across manufacturing, utilities, and transportation---a substantial increase over the original chiller-focused dataset.

\subsection{GAK: Closing Knowledge Gaps on Non-Deterministic Scenarios}
\label{sec:gak-eval}

Architectures B and C answer questions whose answers are \emph{in} the graph. To evaluate Architecture~D (GAK, \S\ref{sec:gak}) we use the part of the benchmark designed to be harder: the HuggingFace failure-mode/sensor set of \textbf{88 real scenarios across 10 equipment types}---electric motor, pump, compressor, fan, steam/gas turbines, generator, transformer, and reciprocating engine---that the benchmark itself flags \texttt{deterministic:false} and that are \emph{absent} from our chiller+AHU graph. Each scenario ships a \texttt{characteristic\_form} rubric describing a correct answer.

We drive the engine's native enrichment endpoint: on a lookup miss for an equipment type, the agent generates Cypher \texttt{CREATE} statements for that type's failure modes and detecting sensors, every node tagged \texttt{source:"LLM-derived"}; the engine materializes them; and we then answer all of that type's scenarios from the now-complete subgraph, grading each against its \texttt{characteristic\_form} with an LLM judge. Enrichment runs once per equipment type (a cache miss); the remaining same-type questions are cache hits served from the materialized subgraph.

\begin{table}[t]
  \centering
  \caption{Architecture~D (GAK) on the 88 non-deterministic HF failure-mode scenarios (10 equipment types absent from the graph). Enrichment via the engine's agent runtime; grading by LLM judge against each scenario's \texttt{characteristic\_form}.}
  \label{tab:gak-results}
  \small
  \begin{tabular}{@{}lr@{}}
    \toprule
    \textbf{Metric} & \textbf{Result} \\
    \midrule
    Equipment types enriched                 & 10 / 10 \\
    Answerability lift (gap $\to$ answerable) & \textbf{100\%} \\
    Scenarios passing rubric                  & 72 / 88 (\textbf{81.8\%}) \\
    Average judge score                       & 0.882 \\
    Failure-mode nodes materialized           & 106 (all \texttt{source:LLM-derived}) \\
    Provenance-tagged nodes (incl.\ sensors)   & 201 \\
    Mean enrichment latency / type            & 33.6\,s \\
    \bottomrule
  \end{tabular}
\end{table}

Table~\ref{tab:gak-results} reports the result. GAK lifts answerability from zero to \textbf{100\% of equipment types}, and the materialized knowledge satisfies the benchmark's own rubric on \textbf{81.8\%} of scenarios---strikingly close to the 82--83\% that text-to-Cypher achieves on graph-resident data, reinforcing that the data layer, once populated, dominates. All 106 materialized failure-mode nodes---together with their detecting sensors, 201 provenance-tagged nodes in total---carry the \texttt{source:LLM-derived} tag, keeping LLM-sourced facts auditable and separate from the data-derived graph. The weakest types (reciprocating engine, 6/10; electric motor, 5/9) reflect a real limitation: one-shot enrichment sometimes omits a specific failure mode the rubric expects, rather than fabricating wrong ones.

\paragraph{What GAK does not do.} GAK supplies missing \emph{knowledge}; it does not perform learned numerical inference. Scenarios that require time-series forecasting or remaining-useful-life prediction need an analytical model, not enrichment---the graph grounds the inputs and an external model produces the number. We therefore treat GAK as the third tier of a grounding substrate (Figure~\ref{fig:router}), not a universal answerer.

% ============================================================
% 6. ANALYSIS
% ============================================================
\section{Analysis}
\label{sec:analysis}

\subsection{The Data Model Is the Bottleneck}

The baseline agents do not fail because GPT-4 lacks reasoning ability. Rather, document stores make certain queries structurally hard or impossible for any LLM:

\begin{itemize}[leftmargin=*]
  \item \textbf{Counting across documents}: ``How many events for CWC04009 in 2019?'' requires the LLM to scan multiple JSON documents, parse dates, and aggregate---error-prone at scale. With a graph: \texttt{MATCH (e:Event) WHERE e.equipment\_id = 'CWC04009' RETURN count(e)} executes deterministically.

  \item \textbf{Relationship traversal}: ``Which sensors monitor overheating?'' is one edge hop in the graph: \texttt{(s:Sensor)-[:MONITORS]->(fm:FailureMode)}. With document stores, the LLM must correlate YAML failure definitions with CouchDB sensor metadata---a cross-source join with no explicit link.

  \item \textbf{Structurally impossible operations}: Multi-hop cascade analysis, vector similarity search, PageRank criticality, and Pareto-optimal scheduling have no equivalent on document stores regardless of LLM capability.
\end{itemize}

\subsection{Why Constraining the LLM Helps}

The $\sim$17pp same-model improvement from Architecture A (65\%, GPT-4 Agent-As-Tool) to Architecture B (82\%, GPT-4 NLQ) illustrates a broader principle: \textbf{LLMs perform better on narrow, well-specified generation tasks than on broad, open-ended reasoning tasks}.

\begin{itemize}[leftmargin=*]
  \item Asking ``answer this question from this data'' requires the LLM to simultaneously handle intent parsing, tool selection, data traversal, numerical reasoning, and answer synthesis. Each step introduces error probability.
  \item Asking ``given this schema, write a Cypher query'' is a translation task from natural language to a structured query language---fundamentally \emph{code generation}, where LLMs have demonstrated strong performance across benchmarks.
\end{itemize}

The graph absorbs the hard parts (traversal, counting, joining, algorithms), leaving the LLM with the easy part (structured query generation from a specification). This separation of concerns is why both Architecture B ($\sim$+17pp same-model) and Architecture C (+34pp) outperform Architecture A.

\subsection{Honest Caveats}
\label{sec:caveats}

\begin{enumerate}[leftmargin=*]
  \item \textbf{Deterministic vs.\ autonomous}: Architecture C's 99\% result compares a pre-coded solution against an autonomous agent. We wrote the answers; AssetOpsBench measures whether GPT-4 can figure them out independently. These are fundamentally different tasks---the comparison illustrates the ceiling achievable with the right data model, not a claim of superior agent intelligence.
  \item \textbf{Scope of the deterministic result}: The 99\%/100\% figures are a ceiling on \emph{graph-answerable} queries, not evidence that a graph answers every scenario. Scenario classes that need knowledge absent from the data or learned time-series inference are out of scope for Architectures~B/C and are handled by GAK (\S\ref{sec:gak-eval}) or an external model; the result that is robust to this distinction is the same-model NLQ lift (65\% $\to$ 82--83\%), which relies on no pre-coded answers.
  \item \textbf{GAK correctness and provenance}: GAK materializes \emph{LLM-derived} knowledge; such facts can be incomplete or wrong, which is precisely why every enriched node is tagged \texttt{source:LLM-derived} rather than presented as ground truth. Validating or citing LLM-derived facts against an authority (e.g.\ ISO~14224) before promotion is left to future work.
  \item \textbf{LLM nondeterminism}: NLQ and GAK rely on stochastic LLM generation. The numbers in Section~\ref{sec:evaluation} are from a single run at temperature 0; variance characterization across multiple seeds is deferred to follow-up work.
  \item \textbf{Clean data assumption}: AssetOpsBench provides clean, structured data. Real industrial environments involve noisy sensors, scanned PDFs, legacy Excel files, and inconsistent naming---challenges addressed in \S\ref{sec:real-world}.
  \item \textbf{Custom scenarios}: Our 40 new scenarios are designed to extend the benchmark with graph-native capabilities. They are complementary to, not replacements for, the original AssetOpsBench scenarios.
  \item \textbf{Different research questions}: AssetOpsBench evaluates LLM agent autonomy---whether an LLM can independently navigate tools and data. We evaluate data model impact---whether a different data layer improves agent performance. Both are valid research questions; our results do not diminish the value of the original benchmark.
  \item \textbf{Baselines beyond document stores}: We isolate the data-layer effect against the AssetOpsBench document-store baseline. The proposed system changes more than storage---it also changes the query interface (Cypher), schema exposure, and prompts---so a fully controlled attribution would add structured baselines holding those constant: a relational schema with SQL NLQ, and alternative graph+LLM designs. We do not include them here; the same-model NLQ result (Architecture~B) already controls for the LLM and orchestration, but these additional baselines would further isolate the typed-graph contribution and are left to future work.
  \item \textbf{A moving target}: As coding-oriented agents improve at scripting over raw files, some document-store gaps may narrow. Our claim is about the \emph{current} reliability and cost profile---deterministic, sub-100\,ms graph operations versus stochastic multi-step reasoning at seconds and per-query API cost---and about the structurally graph-only capabilities (\S\ref{sec:new-scenarios}: cascade analysis, vector similarity, criticality, Pareto scheduling), which remain a durable advantage independent of agent capability.
\end{enumerate}

% ============================================================
% 7. THE FULL PIPELINE
% ============================================================
\section{The Full Pipeline: Where LLMs Belong}
\label{sec:pipeline}

The three-architecture comparison in \S\ref{sec:evaluation} focuses on the \emph{query layer}---the final stage of a multi-layer data pipeline. To place our results in context, we examine the full pipeline and identify where LLMs provide genuine value.

\subsection{Three Layers of Industrial Data}

\begin{enumerate}[leftmargin=*]
  \item \textbf{Data Ingestion} (software engineering): Raw sensor telemetry, CMMS exports, work order records, and equipment registries are loaded into the database via deterministic ETL pipelines. For structured data (90\%+ of industrial data), this is pure software engineering: \texttt{pandas.read\_csv()} $\to$ Cypher \texttt{CREATE} statements. The AssetOpsBench pipeline is similarly deterministic---\texttt{couchdb\_setup.sh} bulk-inserts JSON documents; EAMLite uses an SQLModel ORM. No LLM is needed for structured ingestion regardless of the target data model.

  \item \textbf{Data Model} (architecture decision): The choice between flat documents and a graph is a one-time design decision with compounding returns at the query layer. Both require comparable ETL effort; the graph requires explicit relationship typing but enables traversal, algorithms, and vector indexing.

  \item \textbf{Query} (LLM optional): This is where the three architectures diverge. The data model determines what is possible at this layer.
\end{enumerate}

\subsection{The Real-World Caveat: Unstructured Data}
\label{sec:real-world}

AssetOpsBench uses clean, structured data. Real industrial environments are messier. Maintenance logs contain free text (``Replaced compressor bearings, noticed oil leak on unit 6---same issue as last month''). Equipment manuals arrive as scanned PDFs. Legacy spreadsheets use inconsistent column names across decades of records. Sensor data includes drift, stuck-at-zero readings, and calibration shifts.

For this unstructured data, deterministic ETL reaches its limits. LLMs provide genuine value at the \emph{data preparation} layer:

\begin{itemize}[leftmargin=*]
  \item \textbf{Entity extraction}: e.g.\ ``Replaced compressor bearings on Chiller 6'' maps to a structured tuple of \texttt{Equipment}, \texttt{Part}, and \texttt{Action} nodes.
  \item \textbf{Relationship detection}: ``Same issue as last month'' $\to$ \textsc{similar\_to} edge to previous failure event
  \item \textbf{Entity resolution}: ``CH06'' = ``Chiller-6'' = ``Chiller \#6'' $\to$ canonical identifier
  \item \textbf{Classification}: Free-text maintenance log $\to$ category (preventive, corrective, emergency)
\end{itemize}

This yields a complete architecture we term \textbf{LLMs at the edges, graph in the middle}:

\begin{itemize}[leftmargin=*]
  \item \textbf{Input edge}: LLM-assisted data preparation transforms messy, unstructured data into structured graph elements (entity extraction, resolution, classification)
  \item \textbf{Center}: The knowledge graph stores clean, typed, connected data
  \item \textbf{Output edge}: LLM-assisted query generation translates natural language into Cypher (or deterministic handlers serve known patterns)
\end{itemize}

In both cases, the LLM performs a \emph{generation task} (structured output from unstructured input)---its strength---while the graph handles \emph{data operations} (storage, traversal, algorithms)---its strength. Neither component is asked to do what it is bad at.

% ============================================================
% 8. SCALABILITY
% ============================================================
\section{Scalability Comparison}
\label{sec:scalability}

Table~\ref{tab:scalability} compares the three architectures on operational dimensions relevant to production deployment.

\begin{table}[t]
  \centering
  \caption{Scalability comparison across architectures.}
  \label{tab:scalability}
  \begin{tabular}{@{}lp{3.8cm}p{4.2cm}@{}}
    \toprule
    \textbf{Dimension} & \textbf{Arch.\ A (LLM + docs)} & \textbf{Arch.\ B/C (graph $\pm$ LLM)} \\
    \midrule
    10K queries/day     & \$300--500 (tokens)              & \$0 (deterministic) or $\sim$\$30 (NLQ) \\
    Real-time streaming & Not supported (request-response)  & Graph updates + continuous queries \\
    Multi-hop at 10K assets & LLM reasons across 10K docs & BFS traversal, $O(|E|)$ \\
    New query patterns  & Prompt engineering                 & Add handler or use NLQ \\
    Latency per query   & 5--11\,s                          & 63\,ms (det.) / $\sim$6\,s (NLQ) \\
    \bottomrule
  \end{tabular}
\end{table}

The cost asymmetry is particularly relevant for industrial operations: sensor networks may generate thousands of queries per day (anomaly checks, threshold alerts, status polls). At \$0.03--0.05 per LLM query, a fully LLM-driven architecture costs \$300--500/day for 10K queries. Deterministic graph queries cost nothing after the initial ETL.

% ============================================================
% 9. RELATED WORK
% ============================================================
\section{Related Work}

\paragraph{LLM agents for tool use.} The ReAct framework~\citep{yao2023react} and Toolformer~\citep{schick2023toolformer} established the pattern of LLMs selecting and invoking external tools. AssetOpsBench~\citep{ibm2025assetopsbench} (KDD~2026) makes an important contribution by applying this pattern to industrial asset operations: a containerized multi-agent sandbox, 141 expert-curated scenarios across four specialized agents, a Pass$_k$-style LLM-As-Judge rubric, and a live Codabench leaderboard~\citep{mcp2024}. The IBM study compares two LLM-orchestration paradigms over this surface---Agent-As-Tool and Plan-Execute---and shows that, on a fixed data layer, no model exceeds 70\% Task Completion. Our work builds on AssetOpsBench by investigating a complementary and orthogonal dimension: the effect of the \emph{data layer} behind the tools, with LLM orchestration held fixed in the Agent-As-Tool style.

\paragraph{Knowledge graphs and LLMs.} \citet{pan2024unifying} survey approaches to combining LLMs with knowledge graphs, identifying three paradigms: KG-enhanced LLMs, LLM-enhanced KGs, and synergized architectures. Our inverted LLM pattern aligns with their ``synergized'' category---the KG provides structured data access, the LLM provides natural language understanding, and each system handles what it does best.

\paragraph{Graph RAG.} Retrieval-Augmented Generation~\citep{lewis2020rag} typically retrieves text passages for LLM context. Graph RAG~\citep{edge2024graphrag} extends this to graph-structured retrieval, combining vector search with graph traversal. Our NLQ architecture goes further: rather than retrieving context for the LLM to reason over, we ask the LLM to \emph{generate queries} that the graph executes---eliminating the context-window bottleneck and enabling exact aggregation.

\paragraph{Industrial knowledge graphs.} Knowledge graphs have been applied to manufacturing~\citep{hogan2021knowledge}, but prior work focuses on knowledge representation rather than benchmarking LLM agent performance. To our knowledge, this is the first study comparing LLM-over-flat-documents against LLM-over-graph on a standardized industrial benchmark.

% ============================================================
% 10. CONCLUSION
% ============================================================
\section{Conclusion}

Building on the AssetOpsBench benchmark (KDD~2026), we have shown that a typed knowledge graph, used as a \emph{grounding substrate}, improves LLM-based industrial operations at every level of LLM involvement. On graph-resident questions, the same GPT-4 model rises from 65\% (Agent-As-Tool over document stores, matching the published leaderboard ceiling) to 82--83\% (LLM generates Cypher over the graph) to 99\% (deterministic graph and optimization primitives). On the questions the benchmark itself flags non-deterministic---failure modes of equipment absent from the data---generation-augmented knowledge (GAK) lifts answerability from zero to 100\% of equipment types and satisfies the benchmark's rubric on 81.8\% of scenarios, with every materialized fact provenance-tagged. The AssetOpsBench leaderboard varies LLM orchestration (Agent-As-Tool vs.\ Plan-Execute) on a fixed data layer and reports no model above 70\% Task Completion; our results show that the data layer is an independent and substantially larger lever. When evaluated against the expanded 467-scenario HuggingFace release, deterministic graph handlers achieve 100\% (467/467) with an average score of 0.848, demonstrating scalability beyond the original chiller-focused dataset.

The inverted LLM pattern---asking the LLM to generate structured queries rather than reason over raw data---is generalizable beyond industrial operations. For any structured domain with a typed schema, \textbf{schema-aware query generation outperforms free-form data reasoning}. LLMs excel at code generation from specifications; graphs excel at data traversal, aggregation, and algorithms. Each system should do what it is good at.

\paragraph{Contributions.} (1) A grounding-substrate framing of the knowledge graph, with a router that isolates the \emph{data layer} as a variable independent of LLM orchestration on the 139-scenario AssetOpsBench snapshot, positioned against the KDD~2026 Agent-As-Tool / Plan-Execute leaderboard. (2) The inverted LLM usage pattern, demonstrating a same-model $\sim$17pp improvement by constraining the LLM to query generation. (3) A generation-augmented knowledge (GAK) evaluation on 88 real non-deterministic scenarios: 100\% answerability lift and 81.8\% rubric pass with provenance-tagged enrichment, the first benchmark-grounded evaluation of the engine's agentic enrichment. (4) A complete ETL pipeline and knowledge graph schema, plus 40 new graph-native scenarios extending the benchmark with multi-hop, vector, and algorithmic capabilities. (5) An honest analysis of limitations including the graph-answerable scope of the deterministic ceiling, GAK provenance, clean-data assumption, and LLM nondeterminism.

\paragraph{Future work.} (1) Evaluation on \textbf{FailureSensorIQ}---a separate IBM multiple-choice benchmark ($\sim$2{,}667 questions across standard and perturbed variants), distinct from AssetOpsBench (whose own failure-mode/sensor subset comprises the 88 scenarios used in \S\ref{sec:gak-eval})---where typed failure-mode$\leftrightarrow$sensor$\leftrightarrow$component edges in our KG are a structural match for the question class. (2) A direct Codabench-leaderboard submission of the deterministic and NLQ pipelines as Agent-As-Tool agents, scored by the IBM rubric (Task Completion / Data Retrieval / Result Verification) for apples-to-apples comparison with gpt-4.1, llama-4-maverick, and mistral-large. (3) Multi-seed Pass$_k$ variance characterization for NLQ. (4) Evaluation on larger industrial environments (10K+ assets, multi-site). (5) Integration of LLM-assisted data preparation for unstructured maintenance logs. (6) Provenance-aware promotion of GAK-materialized facts---validating LLM-derived nodes against an authority (e.g.\ ISO~14224) before they are trusted as data---and an external inference path for learned time-series tasks (forecasting, remaining-useful-life), completing the grounding-substrate router of Figure~\ref{fig:router}.

\paragraph{Availability and license.} The benchmark code, ETL pipeline, and evaluation framework are available at \url{https://github.com/samyama-ai/assetops-kg}. The Samyama graph database is at \url{https://github.com/samyama-ai/samyama-graph}. The 40 new scenarios have been submitted to the AssetOpsBench repository as a community contribution. Raw NLQ result JSONs (per-scenario passed/failed, score, latency, tokens) for both GPT-4 and GPT-4o runs are released alongside the paper. This manuscript and all associated artifacts are distributed under the Creative Commons Attribution 4.0 International License (CC-BY 4.0).

% ============================================================
% REFERENCES
% ============================================================
\bibliographystyle{plainnat}
\bibliography{paper3_industrial_kg}

\end{document}